\documentclass[11pt,preprint]{aastex}

\newcommand{\etal}{{\it et~al.}}

\usepackage{longtable}
\begin{document}

\title{Visual-band brightnesses of Near Earth Objects that will be
  discovered in the infrared by NEO Surveyor}

\author{Joseph R. Masiero\altaffilmark{1}, Tyler Linder\altaffilmark{2}, Amy Mainzer\altaffilmark{3}, Dar W. Dahlen\altaffilmark{1}, Yuna G. Kwon\altaffilmark{1}}

\altaffiltext{1}{Caltech/IPAC, 1200 E. California Blvd, MC 100-22, Pasadena, CA 91125 USA, jmasiero@ipac.caltech.edu}
\altaffiltext{2}{Planetary Science Institute, Tucson, AZ 85719 USA}
\altaffiltext{3}{Department of Earth, Planetary and Space Sciences, University of California, Los Angeles, CA 90095 USA}

\begin{abstract}

NEO Surveyor will detect asteroids and comets using mid-infrared
thermal emission, however ground-based followup resources will require
knowledge of the expected visible light brightness in order to plan
characterization observations.  Here we describe the range of
visual-to-infrared colors that the NEOs detected by Surveyor will
span, and demonstrate that for objects that have no previously
reported Visual band observations, estimates of the Johnson
Visual-band brightness based on infrared flux alone will have
significant uncertainty.  Incidental or targeted photometric followup
of objects discovered by Surveyor enables predictions of the fraction
of reflected light visible and near-infrared wavelengths, supporting
additional detailed characterization.

\end{abstract}

\section{Introduction}

The Near-Earth Object (NEO) Surveyor mission will survey the Solar
system at two thermal infrared wavelengths ($4-5.2~\mu$m and
$6-10~\mu$m) with the singular goal of detecting the majority of
objects that pose a regional hazard to Earth.  Over the course of the 5-year
required survey period, the mission is expected to detect
$\sim100,000-200,000$ near-earth asteroids larger than $25~$m in size,
and will report these observations to the MPC within 3 days of data
acquisition.  \citet{mainzer23} provide an overview of the mission
and the design of survey simulation tools that will be used for survey
completeness validation.

The survey strategy employed by the NEO Surveyor mission is designed
to provide sufficient self-followup to establish the orbit and
effective spherical diameter of the NEOs observed. Diameters can be
determined from NEO Surveyor data even if visible light flux is not
available \citep{mainzer15}, as was done for the objects discovered by
the NEOWISE survey that did not receive ground based followup
\citep[e.g.][]{mainzer11neo,masiero11}. This is possible because the
total thermal emission is only weakly dependent on albedo, and so the
added uncertainty due to the unknown albedo is only a small component
of the total error.

The mission also has a Targeted Followup Observation (TFO) mode for
high-priority objects that can be used for improving orbital
knowledge, obtaining high-cadence light curve observations (individual
\emph{Exposures} are taken every 30 s), and improving thermal modeling
constraints.  The TFO mode, however, takes time away from normal
survey operations and so will only be used sparingly for the most
critical objects.  Additional characterization such as spectroscopy
would need to employ ground- or space-based followup telescopes, the
vast majority of which observe at visible or near-infrared
wavelengths.  In order to support asteroid and comet followup
observations by the community, the NEO Surveyor mission will provide
estimated Johnson Visual magnitude brightnesses for discovered objects
to enable these investigations.

The recent International Asteroid Warning Network
(IAWN\footnote{\textit{https://iawn.net/}}) observing campaigns
present a model for the types of followup that could be obtained for a
potential impactor.  IAWN is a global collection of amateur and
professional astronomers dedicated to studying and monitoring the
near-Earth asteroid population. IAWN to date has held six asteroid
campaigns, two of which were dedicated to the observational timing
assessments of fast-moving asteroids and four were intended to assess
the ability of the community to rapidly recover and characterize NEOs.
Those latter four campaigns in particular demonstrated how additional
characterization data would be used to help assess the hazard posed by
an impactor, and how this knowledge would be disseminated among the
various stakeholders. The types of characterization data that could
possibly be obtained and the timing for when it would be available can
have significant impacts on the responses that are available.
Therefore, it is important to understand the resources that would be
able to access new NEO discoveries.

The most recent campaign was a rapid response exercise following the
discovery of NEO 2023 DZ2, whose impact probability quickly rose with
follow-up observations before measurements from precovery data retired
the impact risk \citep{reddy24}. This object was discovered at a
$V=20.3~$mag ten days prior to close approach, and rose to a peak
brightness of $V\sim13~$mag, allowing a wide range of characterization
tools to be used. Visible and near-infrared photometry, spectroscopy,
and polarimetry all provided an immediate estimate of the albedo of
this NEO that was fed into the hazard model.

Previous characterization campaigns focused on recovery and tracking
of 2012 TC4 \citep{reddy19}, spectral characterization of (66391)
Moshup \citep{reddy22kw4}, and simulated discovery and hazard
assessment of (99942) Apophis \citep{reddy22apophis}.  The
characterization campaign for 2012 TC4 was able to obtain light curve
properties while the object was at $V\sim22.5~$mag, however these
rotational observations were at the practical limit for the
$4-5~$m-class telescopes used. In each case, ground-based followup
allowed the impact hazard model to be refined and improved.

To assess the potential future capacity for followup of NEO Surveyor
discoveries, we take the synthetic population of near-Earth objects
that were created to verify the mission's Level 1 survey requirements
and extend this population down to diameters of $D=25~$m.  This
population is assigned both orbital and physical properties.  Using
the current best knowledge of the survey rules and system
sensitivities for NEO Surveyor, we determine the Visual and
mid-infrared fluxes of each NEO at the time when it would first be
discovered by the survey.  We present the results of this analysis and
related discussion below.

\section{Population Model and Survey Simulation}

We use the survey simulation tools recently developed for the NEO
Surveyor mission to generate a synthetic population of near-Earth
objects down to sizes below the current completeness limit.  We
simulate objects with diameters of $D\ge25~$m following the
size-frequency distribution described in \citet{mainzer23}, with
orbital and albedo distributions based on the nearly-complete regime
of the currently known NEO population \citep[see ][for more details
  about the NEO size distribution]{neosdt}.  Our survey simulator
calculates the position and expected flux of each object for every
time it would be in the field of view of a survey pointing.  Dynamical
and flux calculation routines were validated against standard software
tools as well as the NEOWISE data, as described in \citet{masiero23}.

Predicted fluxes for NEOs are compared to the current best estimate of
the system sensitivity and the predicted zodiacal light background
flux for each observation to determine if the synthetic object would
be detected.  A fraction of these potential detections are then
discarded based on a signal-to-noise-dependent factor to account for
incompleteness in the data extraction due to e.g. confusion, bad
pixels, etc.  The remaining detections are then assembled into
tracklets following the same rules that will be used when processing
the flight data.  This produces a set of data analogous to what the
NEO Surveyor will be submitting to the Minor Planet Center once the
survey commences.

Because physical properties are generated for each object, it is
possible to determine the Visual band flux from reflected light as
well as the thermal infrared flux from emission at the time each
object would be detected.  While thermal emission is determined using
the Near-Earth Asteroid Thermal Model \citep[NEATM][]{harris98},
Visual band flux is determined using the H-G asteroid photometric
system \citep{bowell89} and the orbital geometry at the time of
observation.  Using these two fluxes for each detection, we can
determine the apparent magnitude difference from V band to the NC2
bandpass centered at $8~\mu$m (the band where Surveyor will be most
sensitive to NEOs).  In this implementation, this difference doesn't
strictly indicate a spectral reflectance gradient, however we will
still refer to it here as the $V-NC2$ color.  

Approximately $40\%$ of the NEOs larger than $140~$m in diameter are
expected to be known at the time of launch based on current survey
capabilities \citep{grav23} and thus will already have visible light
measurements available to use for estimating visible brightnesses.
The Vera Rubin telescope is scheduled to begin survey prior to the
launch of NEO Surveyor, and so will improve the catalog of reflected
light measurements with observations of a large number of Solar system
objects. \citet{wagg24} show that Rubin will typically observe NEOs at
$V\sim22-23$ (see their Figure 7), but will observe few NEOs with
$V>24~$mag. The remaining NEOs detected by Surveyor will have unknown
visible light properties (in particular an unknown Visual band albedo)
and as such would be expected to be drawn randomly from the population
below.  Quantification of this uncertainty is needed to properly plan
for any followup observations.

\section{Survey Simulation Results}

\begin{figure}[ht]
\begin{center}
  \includegraphics[scale=0.7]{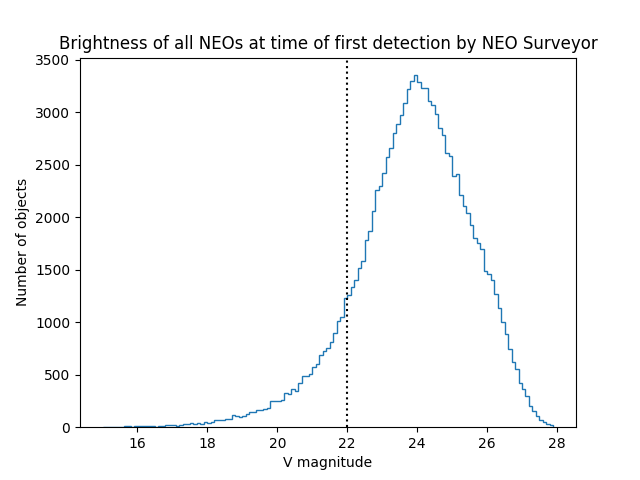}
  \protect\caption{V-band magnitude distribution of all simulated NEOs
    at the time they would be first detected by NEO Surveyor.  The
    vertical dotted line indicates V$=22~$mag, the approximate
    limiting magnitude of current NEO surveys and followup.  }
\label{fig.vhist}
\end{center}
\end{figure}

We show in Figure~\ref{fig.vhist} the distribution of Johnson V-band
magnitudes for all NEOs at the first time that they would be detected
during the survey.  We use the time of the initial detection, rather
than the peak brightness of each object over the 5-year survey period,
as it is more representative of the brightness at time when immediate
characterization followup would be most desirable for an object with
non-zero impact probability, and when the predicted on-sky position
will be best known prior to followup.  We mark the location of
$V=22~$mag in this plot to show the approximate limits for the most
common NEO survey and followup assets in use today
\citep[e.g.][]{larson03,denneau13,bellm19,devogele19}.  As
demonstrated in this plot, the vast majority of NEOs detected by
Surveyor will be magnitudes fainter than what is currently submitted
to the MPC and published for followup on the NEO Confirmation
Page\footnote{\it{https://minorplanetcenter.net/iau/NEO/toconfirm\_tabular.html}}.
This is consistent with the previous findings of \citet{mainzer15}.

\begin{figure}[ht]
\begin{center}
  \includegraphics[scale=0.7]{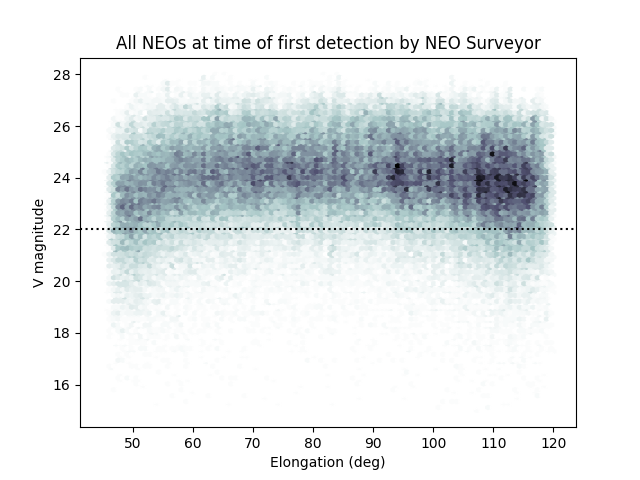}  
  \protect\caption{V-band magnitude as a function of elongation for
    all simulated NEOs at the time they would be first detected by NEO
    Surveyor.  The horizontal dotted line indicates V$=22~$mag, the
    approximate limiting magnitude of current NEO surveys and
    followup.  Spatial bins in this plot represent a linear count of objects.}
\label{fig.v_elong}
\end{center}
\end{figure}

Figure~\ref{fig.v_elong} shows the spatial histogram of NEO V
magnitudes compared to the elongation (on-sky angular distance from
the Sun) at the time of first detection by NEO Surveyor.  The
magnitude distribution is consistent across most elongations, with a
$\sim1~$mag brightening at the lowest elongations.  However, low
elongations are the hardest for ground-based telescopes to access as
they are only above 2 airmasses in the twilight sky or for a short
period of time outside twilight.  Incidental visible light photometry
will likely be available for objects at larger elongations from
ground-based surveys such as the Vera C. Rubin telescope
\citep{jones18}.  The trend of increasing numbers of objects at larger
elongations is due primarily to the large population of amor-class
NEOs, which spend all their orbit outside the Earth's, and thus are
most accessible at high elongations. Amors tend to have higher minimum
orbital intersection distances (MOID) with the Earth's orbit, and thus
are less likely to be categorized as potentially hazardous objects
which are defined as objects with MOID$<0.05~$AU.

\begin{figure}[ht]
\begin{center}
  \includegraphics[scale=0.7]{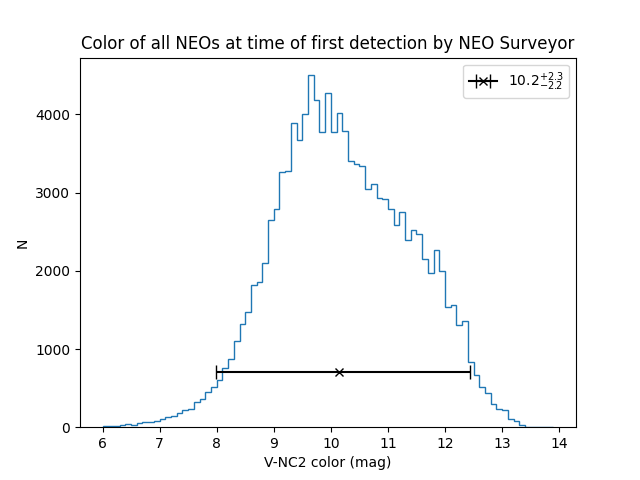} \protect\caption{V-NC2
    color distribution of all simulated NEOs at the time they would be
    first detected by NEO Surveyor. The cross indicates the median
    value of the distribution, while the whiskers show the range
    spanned by the $2.5^{\rm{th}}$ and $97.5^{\rm{th}}$ percentiles.}
\label{fig.colorhist}
\end{center}
\end{figure}

The planning of any desired characterization followup of NEO Surveyor
discoveries beyond mission-obtained TFO characterization will be
difficult for objects that do not receive any incidental visible light
observations, and thus have unknown brightnesses at reflected light
wavelengths.  Because of the wide range of albedos that have been
observed for NEOs \citep{mainzer11neo,wright16} and the albedo
distribution that is expected for the population observed by NEO
Surveyor \citep{mainzer23}, the predicted reflected light brightness
for an object of a given infrared flux will be drawn from a large
potential range.  Figure~\ref{fig.colorhist} shows the distribution of
V-NC2 colors for all NEOs in our simulation at the time of first
detection.  While the distribution is peaked at $V-NC2\sim10$, the
full-width at half-maximum of this distribution spans
$\sim3~$mags. The figure also shows the median $V-NC2$ color (10.2) as
well as the $2.5^{\rm{th}}$ and $97.5^{\rm{th}}$ percentile ranges
($-2.2$ and $+2.3$, respectively.) which span the 2-sigma region of
the distribution.

\begin{figure}[ht]
\begin{center}
  \includegraphics[scale=0.7]{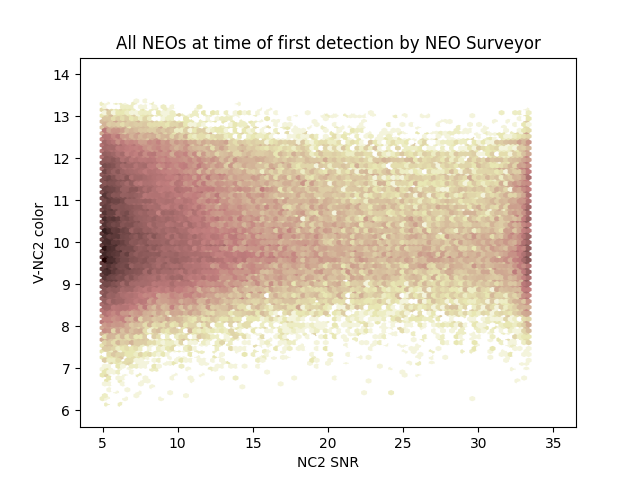}
  \protect\caption{V-NC2 color vs signal-to-noise (SNR) of the NC2
    detection for simulated NEOs at the time of first
    detection. Spatial bins in this plot represent a logarithmic count
    of objects; the vast majority of detections are at SNR$\sim5$.
    Due to the expected photometric systematic uncertainties, sources
    have an effective upper limit to their SNR causing the apparent
    increase in source density at SNR$\sim34$ for all objects brighter
    than this limit.}
\label{fig.color_snr}
\end{center}
\end{figure}

\begin{figure}[ht]
\begin{center}
  \includegraphics[scale=0.7]{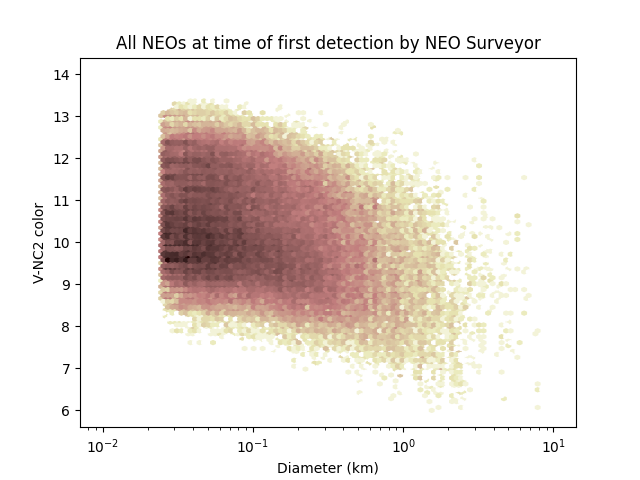}
  \protect\caption{V-NC2 color at the time of first detection vs the
    size of the simulated NEOs. Spatial bins in this plot represent a
    logarithmic count of objects.  Larger objects are skewed bluer as
    they are more likely to be detected further from the Sun, but they
    still span a significant range of colors.}
\label{fig.color_diam}
\end{center}
\end{figure}

Other detection parameters available at the time of observation, such
as elongation or signal-to-noise in the NEO Surveyor bandpasses, do
not provide constraints on the expected $V-NC2$ color.  As shown in
Figures~\ref{fig.color_snr}, the color distribution does not show
significant structure with respect to brightness at the NC2
wavelengths.  Parameters that can be derived from NEO Surveyor data
after self-followup and orbit fitting such as diameter also show large
spreads in $V-NC2$ color, as shown in Figure~\ref{fig.color_diam}.
Larger objects do tend to show bluer colors, as they are more likely
to be detected at larger heliocentric distances where their
temperatures are lower, however these are also the most likely to be
already recorded in the current catalog.

\section{Discussion}

Our simulations show that there is a wide dispersion in possible $V$
band magnitudes for infrared-detected NEOs.  It is not possible to
make a precise prediction of these objects' expected Visual magnitudes
using only the data obtained by NEO Surveyor.  This means that
targeted followup of individual objects of interest will need to
account for a wide range of possible brightnesses, where only part of
the possible range may be detectable with the followup instrumentation
available, e.g. spectroscopy.

A complete analysis of the capabilities of follow-up assets currently
available is challenging due to the large number of variables. The
location of the telescope, observing time availability, and specific
instrument parameters such as image scale and readout time make a
direct general comparison difficult.  Instead, we discuss a few
specific cases commonly needed for astrometric followup.

The most critical type of astrometric observation will be if NEO
Surveyor discovers a virtual impactor. Selecting the correct
ground-based assets is crucial to successful recovery. For example,
the first IAWN campaign recovered 2012 TC4 using VLT at $V\sim27~$mag
which is the practical upper limit of recovery using an 8 meter
ground-based telescope.  The NEO community will likely be limited by
observing time constraints to a handful of recoveries at these
brightnesses per telescope per year. Considering all the 8 meter
telescope time that could be even potentially be available for
recovery, less than 50 asteroids per year could receive astrometric
follow-up fainter than $V=26~$mag.

The typical limiting magnitude of 2-4m class telescopes is
$V\sim25~$mag. A 2 meter-class telescope optimized for moving object
astrometric recovery can outperform a 4 meter telescopes with a more
generalized setup and instrument set.  However, the availability of
facilities for asteroid recovery between $V\sim23-25~$mag will be
limited to a theoretical maximum of hundreds of asteroids per night if
all potential facilities were used in a dedicated fashion. Objects
brighter than $V\sim23~$mag are in the range of the current NEO survey and
followup programs, and so could be expected to receive followup based
on the current operational program, however the number of potential
targets will likely saturate the available time for the currently
operational programs. 

Photometric and spectroscopic characterization observations are
limited to $V\sim22~$mag currently, which means that it will be
difficult to obtain further analysis for many members of the
population discovered by NEO Surveyor.  This is especially true for
objects where the possible Visual band brightness is predicted to
straddle this sensitivity threshold.  One method to address this
challenge is for ground-based telescopes to report negative detections
and the limiting magnitude that was searched. This would allow for
setting a limit on the possible $H_V$ magnitude and rule out the
possibility of characterization for some objects.

It is important for the NEO Surveyor project to report the minor
planet observations in a way that provides as much utility as possible
for the followup observer community.  Experience from NEOWISE, where
the team reported estimated $V$ magnitudes for most objects to aid in
followup, found that predicted $V$ magnitudes could often be $2~$mags
fainter than the recovery magnitude, as NEOWISE preferentially
discovered low-albedo NEOs \citep{mainzer11neo}.  Some
NEOWISE-discovered objects had no reported estimated visible
brightness at all, resulting in objects in the orbit catalog without
$H_V$ magnitudes, which significantly hampered followup planning.  The
NEOWISE reported $V$ magnitudes were estimates based on the infrared
flux, not visible light measurements, but this fact is not explicitly
indicated as such in the MPC observation archive.  Users of the data
would need to refer to the NEOWISE documentation
\citep[e.g.][]{cutri12,cutri15} to learn the source of these values.

To resolve this situation, we propose to use the new reporting fields
enabled by the Astrometry Data Exchange Standard
(ADES\footnote{\textit{https://minorplanetcenter.net/iau/info/ADES.html}})
format, the preferred format for submitting data to the Minor Planet
Center.  Specifically, the mission can report estimated $V$ magnitudes
as was done for NEOWISE, but now can include an uncertainty on that
value that will indicate it should only be considered a weak
constraint on the actual expected $V$ mag.  Based on the distribution
shown in Fig~\ref{fig.colorhist}, and the range spanned by the 2-sigma
region, the value to report would be
\[V_{estimate} \approx NC2 + 10 \pm 2~\rm{mag}\]
which would cover the majority of the expected $V~$mag distribution.

The benefit of a large uncertainty value is that any real photometric
measurement that follows the NEO Surveyor observations would quickly
dominate the $H_V$ calculation carried out during orbit fitting
(assuming magnitude uncertainties are also reported for those
measurements).  Expansion of the MPC's NEO Confirmation Page to
include the uncertainty on both the predicted $V$ and $H_V$ values
will enable observers to understand the provenance of these predicted
values.

Beyond NEO Surveyor, estimated $V$ magnitudes may be useful for
observation reports from other infrared observatories such as the
James Webb Space Telescope (JWST). As described by \citet{muller23},
during calibration observations the JWST Mid-Infrared Instrument
(MIRI) detected a previously unknown asteroid, and the authors
estimate that every MIRI observation near the ecliptic will likely
contain a few unknown asteroids.  These MIRI data alone were
sufficient to constrain the size of this unknown object, despite the
lack of orbital arc coverage needed to fit an orbit.  While
astrometric recovery and followup of these short-arc objects is nearly
impossible with current resources, the ability to report visual
magnitude estimates for infrared detections will potentially help
future efforts to link these detections to later observations.

\section{Conclusions}

We use the survey simulation tools developed for NEO Surveyor to
investigate the expected brightnesses of near-Earth objects that the
mission will discover.  We find that the vast majority of objects will
be significantly fainter than the current depth reachable with most
reflected light followup assets.  Ground-based targeted astrometric
followup will be difficult to achieve on a large scale, although the
mission's survey is designed to provide astrometric self-followup.  In
order to obtain optical photometry and spectral characterization of
any virtual impactors discovered by NEO Surveyor it will be necessary
to invest in more sensitive followup resources and exercise these
systems.

Estimating the Visual brightness of NEO Surveyor discoveries is
complicated by the wide range of possible $V-NC2$ colors that detected
objects are predicted to have.  Reports of these estimates will use
the photometric uncertainty field to identify to followup observers
the large range of brightnesses that are possible.  Propagation of
these uncertainties to the predicted apparent and absolute magnitudes
provided on the MPC's NEO Confirmation Page will allow followup
observers to plan for a range of activities based on the recovered
brightness.  This will fulfill the needs of the followup community for
observation planning purposes, as well as avoid empty entries in the
orbit catalogs, all while making it clear to data users that this
value is only an estimate of the actual brightness of the object.

\section*{Acknowledgments}
We thank the referees for their helpful comments that improved this
manuscript. This publication makes use of software and data products
from the NEO Surveyor, which is a joint project of the University of
California, Los Angeles and the Jet Propulsion Laboratory/California
Institute of Technology, funded by the National Aeronautics and Space
Administration.  This research has made use of data and services
provided by the International Astronomical Union's Minor Planet
Center.  This research has made use of NASA’s Astrophysics Data System
Bibliographic Services.  This research has made use of the {\it
  neospy}, {\it numpy}, {\it scipy}, {\it astropy}, and {\it
  matplotlib} Python packages.

\end{document}